\documentclass[final,5p,times, twocolumn]{elsarticle}

\usepackage{amssymb}
\usepackage[numbers]{natbib}
\usepackage[most]{tcolorbox}
\usepackage{dirtytalk}
\usepackage{fontawesome}
\usepackage{pifont}
\usepackage{balance}
\usepackage[T1]{fontenc}
\usepackage{ragged2e}
\usepackage{makeidx}
\usepackage{multicol}
\usepackage{multirow}
\usepackage{lipsum}
\usepackage{footnote}
\usepackage{diagbox}
\usepackage{longtable}
\usepackage{tabularx}
\usepackage{colortbl}
\usepackage{hyperref}
\usepackage{footmisc}
\usepackage{float}
\usepackage{amsthm}
\usepackage{graphicx}
\usepackage{pythonhighlight}
\usepackage[figuresright]{rotating}
\usepackage{amsmath}

\definecolor{anti-flashwhite}{rgb}{0.95, 0.95, 0.96}
\definecolor{beige}{rgb}{0.96, 0.96, 0.86}
\definecolor{floralwhite}{rgb}{1.0, 0.98, 0.94}
\definecolor{gainsboro}{rgb}{0.86, 0.86, 0.86}
\definecolor{ghostwhite}{rgb}{0.97, 0.97, 1.0}
\definecolor{honeydew}{rgb}{0.94, 1.0, 0.94}
\definecolor{isabelline}{rgb}{0.96, 0.94, 0.93}
\definecolor{ivory}{rgb}{1.0, 1.0, 0.94}
\definecolor{magnolia}{rgb}{0.97, 0.96, 1.0}
\definecolor{mintcream}{rgb}{0.96, 1.0, 0.98}
\definecolor{pearl}{rgb}{0.94, 0.92, 0.84}
\definecolor{whitesmoke}{rgb}{0.90, 0.90, 0.90}

\def\tsc#1{\csdef{#1}{\textsc{\lowercase{#1}}\xspace}}
\tsc{WGM}
\tsc{QE}
\tsc{EP}
\tsc{PMS}
\tsc{BEC}
\tsc{DE}

\begin{document}

\begin{frontmatter}

\let\WriteBookmarks\relax
\def\floatpagepagefraction{1}
\def\textpagefraction{.001}

\title{\textsc{RoseMatcher}: Identifying the Impact of User Reviews on App Updates}

\author[1,2]{Tianyang Liu}
\ead{til040@ucsd.edu}
\address[1]{School of Computer Science, Wuhan University, 430072 Wuhan, Hubei, China}
\address[2]{Department of Computer Science and Engineering, University of California San Diego, La Jolla, California 92093, United States}

\author[1,3]{Chong Wang\corref{cor1}}
\ead{cwang@whu.edu.cn}

\author[1]{Kun Huang}
\ead{hunk_fe@whu.edu.cn}

\author[1]{Peng Liang\corref{cor1}}
\ead{liangp@whu.edu.cn}

\author[1]{Beiqi Zhang}
\ead{zhangbeiqi@whu.edu.cn}

\author[3]{Maya Daneva}
\ead{m.daneva@utwente.nl}
\address[3]{Faculty of Electrical Engineering, Mathematics and Computer Science, University of Twente, 7500 AE Enschede, The Netherlands}

\author[3]{Marten van Sinderen}
\ead{m.j.vansinderen@utwente.nl}



\cortext[cor1]{Corresponding author}

\begin{abstract}
\textbf{Context}: The release planning of mobile apps has become an area of active research, with most studies centering on app analysis through release notes in the Apple App Store and tracking user reviews via issue trackers. However, the correlation between these release notes and user reviews in App Store remains understudied.

\noindent\textbf{Objective}: In this paper, we introduce \textsc{RoseMatcher}, a novel automatic approach to match relevant user reviews with app release notes and identify matched pairs with high confidence.

\noindent\textbf{Methods}: We collected 944 release notes and 1,046,862 user reviews from 5 mobile apps in the Apple App Store as research data to evaluate the effectiveness and accuracy of \textsc{RoseMatcher}, and conducted deep content analysis on matched pairs.

\noindent\textbf{Results}: Our evaluation shows that \textsc{RoseMatcher} can reach a hit ratio of 0.718 for identifying relevant matched pairs, and with the manual labeling and content analysis of 984 relevant pairs, we identify 8 roles that user reviews play in app updates according to the relationship between release notes and user reviews in the relevant matched pairs.

\noindent\textbf{Conclusions}: Our findings indicate that both app development teams and users pay close attention to release notes and user reviews, with release notes typically addressing feature requests, bug reports, and complaints, and user reviews offering positive, negative, and constructive feedback. Overall, the study highlights the importance of the communication between app development teams and users in the release planning of mobile apps, with relevant reviews tending to be posed within a short period before and after the release of release notes, with the average time interval between the post time of release notes and user reviews being approximately one year.
\end{abstract}


\end{frontmatter}



\section{Introduction}  
\label{introduction} 
With the rapid progress in mobile techniques and smartphones, the number of mobile applications (apps for short) has risen dramatically in recent years. As of the second quarter of 2022, Android users were able to choose between 3.5 million apps, making Google Play the app store with the biggest number of available apps. The Apple App Store is the second-largest app store with roughly 2.2 million available apps for iOS~\cite{iOSStat}. App stores now become the primary data source to construct app datasets for the research and practice on app development, evolution, and maintenance \cite{AppStoreSurvey}.

\begin{figure*}[htbp]
\centering
  \includegraphics[width=\textwidth]{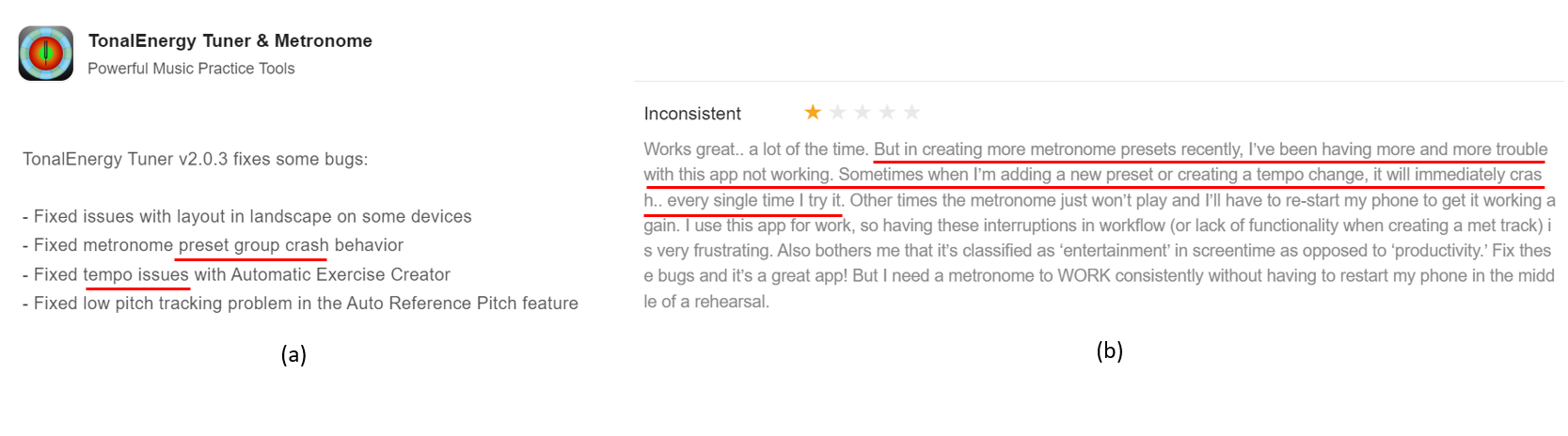}\\
  \caption{Exemplary release note (a) and user review (b) from app TonalEnergy Tuner \& Metronome in Apple App Store.}
  \label{fig:1.1}
\end{figure*}

In these app stores, app vendors regularly deliver official release notes with the new releases of their apps to highlight the key or essential updates of its current version, as shown in Figure~\ref{fig:1.1} (a). Meanwhile, users freely post their reviews on the apps they use in the app store, including praises, complaints, feature requests, and recommendations \cite{UserFeedback}, as shown in Figure~\ref{fig:1.1} (b). Several existing studies have reported that it is essential for app vendors to regularly release new versions to fix bugs or introduce new features \cite{FreshApps}, since user dissatisfaction can quickly lead to the fall of even popular apps \cite{Satisfaction}. This implies that constantly monitoring and capturing, and proactively meeting user needs could lead to a successful app. For example, as the underlined text in Figure~\ref{fig:1.1} shows, a user complained about the crash when he/she added a new preset or created a tempo change (preset and tempo are terms used in the music field), and this bug was fixed in the subsequent release of this app. 

This circumstance shows that user reviews could be the potential evolutionary requirements for for app development teams to update apps, and write the updating details in release notes for users.

Although lots of users feedback is generated in different platforms, such as Apple App Store~\cite{UserFeedback} or social media~\cite{Twitter}, only 35.1\% user reviews were reported to contain valuable information for app development teams~\cite{ARMiner}. Manual extraction of key valuable information from user feedback would inevitably be time-consuming, laborious, and clumsy. In recent years, many researchers studied the filtering of informative user reviews~\cite{ARMiner}, the identification of relevant user feedback~\cite{AnalysisOfUserComments}, and clustering and prioritization of user requirements in user reviews~\cite{ARMiner, CLAP}. These studies proposed various methods to extract critical information for app evolution from large amounts of data, but no analysis was conducted on those user reviews from the perspective of app release notes, i.e., to explore the echoing relationship between user reviews and release notes.

To this end, we propose \textsc{RoseMatcher} (\textbf{R}elease n\textbf{O}te and u\textbf{S}er r\textbf{E}view \textbf{Matcher}), which is a novel approach that matches colloquially written user review sentences with formally written release note sentences. With \textsc{RoseMatcher}, we can obtain high-confidence matched pairs of relevant user reviews and release notes, thus significantly reducing the manual annotation effort. Based on these matched pairs, we detect an echo relationship between app release notes and user reviews, and we explore the roles of user reviews in app updates from this mutual response and conduct analysis at the level of their posting time.

To the best of our knowledge, seldom research explores the relevance between user reviews and app release notes. 
In this paper, we focus on those user reviews that are deemed to be relevant to specific release notes. First, we defined eight roles user reviews take in app updates based on the content association between release notes and user reviews. Subsequently, we explored the attention app vendors pay to user reviews and the attention users pay to release notes in terms of the temporal dimension, based on the difference between the release time of release notes and the post time of user reviews.

The main contributions of this study are listed below:

\begin{itemize}

    \item To detect the relevance between official release notes and informal user reviews, we propose \textsc{RoseMatcher} to mine the most relevant reviews for app release notes by combining semantic-sensitive and keyword-based matching algorithms. \textsc{RoseMatcher} can detect semantically similar and keyword-identical texts and solve the problem of the low relevance rate of high-similarity matched pairs suggested by the single model, which dramatically reduces the overhead and cost of manual annotation.
    \item To explore the roles user reviews played in app updates, we manually analyze and compare the content of the relevant user reviews matched by \textsc{RoseMatcher} and their relevance with release notes, and finally, we define eight different roles user reviews play in app updates.
    \item To explore the attentiveness of the app development teams to user reviews and users' attentiveness to app release notes posted in the app market, we analyze and visualize the time interval between the release time of the release notes and the post time of their relevant user reviews. To our best knowledge, no prior study analyzed release notes and their relevant user reviews in the temporal dimension, and our study fills the gap in this research area.

\end{itemize}

The rest of the paper is organized as follows. Section \ref{rm} elaborates our proposed methodology for identifying relevant matched pairs between user reviews and app release notes, including overview in Section~\ref{overview}, data pre-processing and selection in Section~\ref{dfp}, algorithms for matched pair identification in Section~\ref{rose}, and specific methodology for matched pair analysis in Section~\ref{ra}. 
 
Section~\ref{er} presents the data collection and experimental results. Section \ref{discussion} elaborates the corresponding interpretation derived from the experimental results, and Section~\ref{implication} explicitly states the implications from different perspectives. Limitations are explicitly given in Section~\ref{tv}. In Section ~\ref{rw}, we briefly review the related work on user reviews and app release notes.  Finally, Section \ref{cfw} concludes the paper and describes the possible future work of our study.

\section{Research Methodology}
\label{rm}

\subsection{Overview of our Approach}
\label{overview}

\begin{figure*}[htbp]
  \centerline{\includegraphics[width=0.9\textwidth]{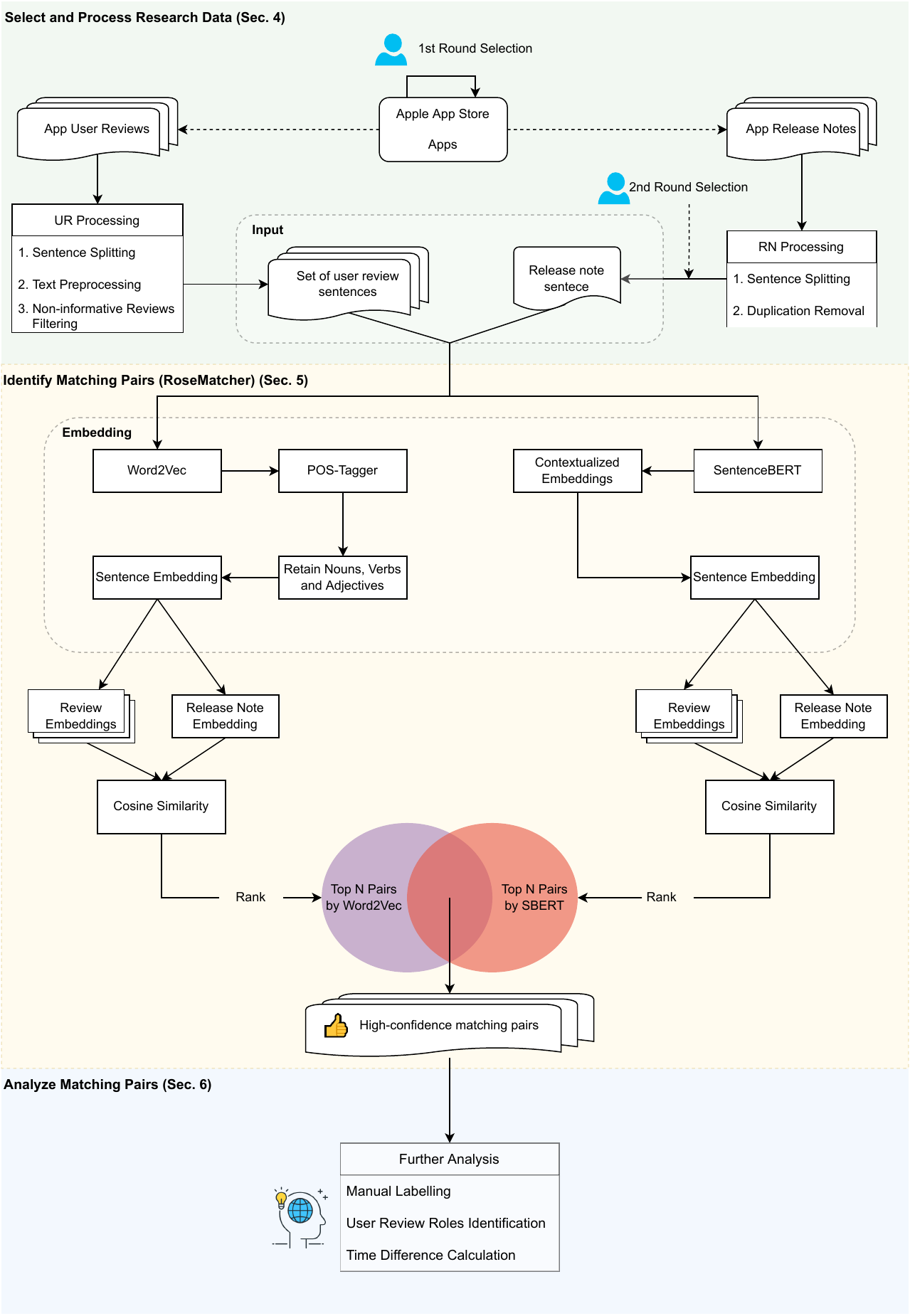}}
  \caption{Overview of the data processing and \textsc{RoseMatcher} approach.}
  \label{fig:overview}
\end{figure*}

In our approach, researchers can get high-confidence matched pairs of release note sentences and user review sentences for a more in-depth study, this process mainly contains the following three steps as shown in Figure~\ref{fig:overview}.

\textbf{(1) Data Selection and Processing.} focuses on data selection, pre-processing, and filtering of the acquired dataset to ensure a high quality of data input for \textsc{RoseMatcher} and thus prevent invalid calculation, which is elaborated in Section~\ref{dfp}.

\textbf{(2) Matched Pairs Identification (\textsc{RoseMatcher}).} The second step is the core of our entire approach -- the \textsc{RoseMatcher}. As described in Section \ref{rose}, we use deep learning models to encode sentence-level release notes and user reviews, in order to get matched pairs by sentence similarity calculation and ranking.

\textbf{(3) Matched Pairs Analysis.} The third step is the subsequent analysis of the suggested high-confidence matched pairs. In Section~\ref{ra}, we elaborate on our specific methodologies for analyzing matched pairs.

\subsection{Data Selection and Processing}
\label{dfp}
Due to the huge amount of computation, we need to make sure we have high-quality input data, which can heavily reduce unnecessary computational overhead. The data selection, filtering, and processing is to ensure the quality of our data as much as possible before they are input to \textsc{RoseMatcher}, thus making our algorithm more efficient.

\subsubsection{Selection Criteria of Mobile Apps}
\label{as}

\begin{figure*}[t]
  \includegraphics[width=\textwidth]{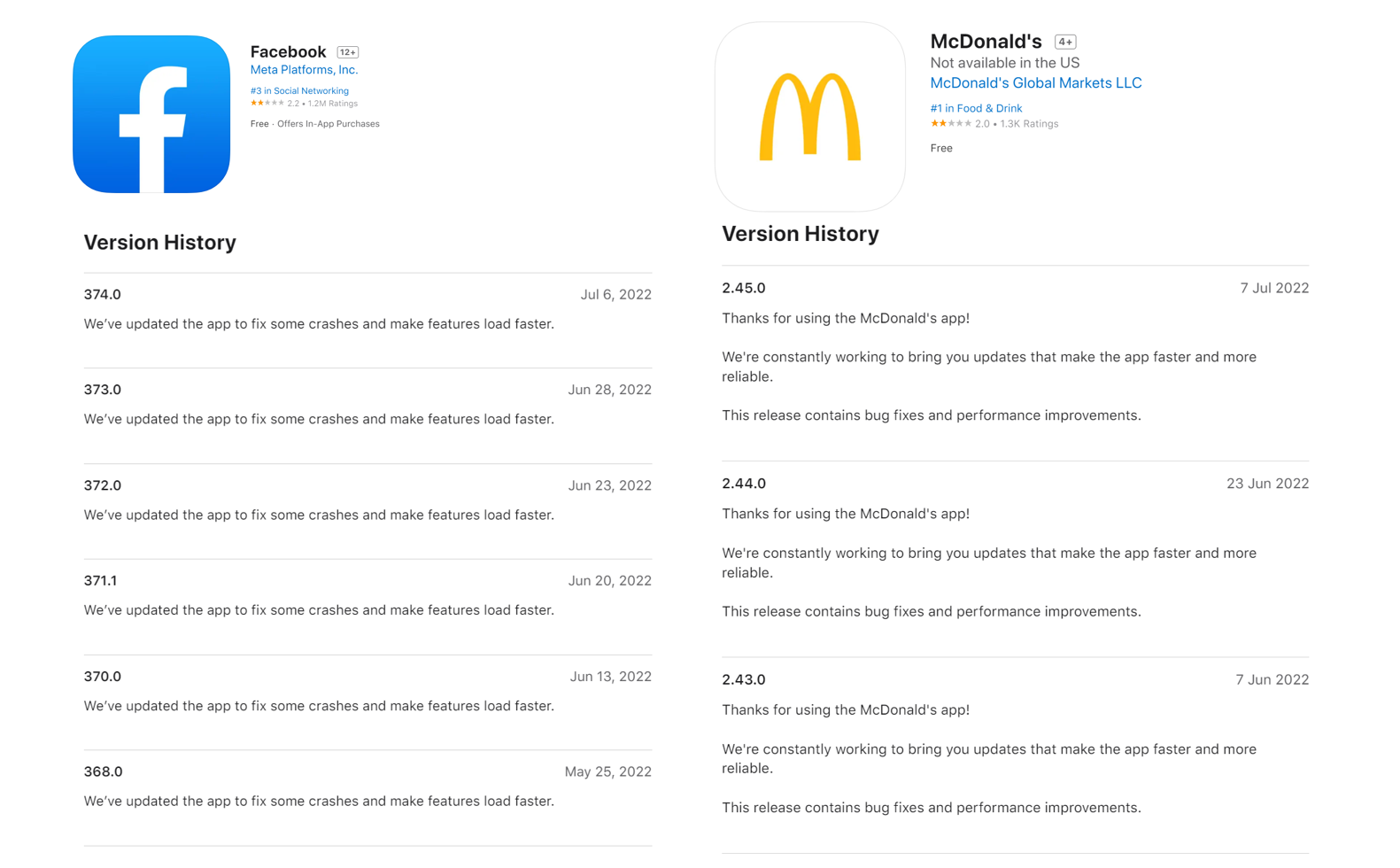}\\
  \centering
  \caption{Examples of repetitive app release notes that convey valueless information.}
  \label{fig:bad update}
\end{figure*}

As reported in~\cite{AnEmpirical}, nearly 60\% of the mobile apps in app stores are updated in a non-updating pattern, i.e., the same release notes are pushed repeatedly. Meanwhile, we observed that some apps usually release notes in a perfunctory manner with broad and repetitive statements, rather than 
specific updating details.
As shown in Figure~\ref{fig:bad update}, for example, the release notes of \textit{Facebook} and \textit{Mcdonald's} are deemed to be low-qualified because they seldom convey specific or valuable information for RE researchers. 
Therefore, we introduced the following four inclusion criteria (IC) to select mobile apps with highly qualified release notes. Note that in this paper, it works as the first-round selection for the construction of our research dataset.

\begin{itemize}
    \item[IC1.1] The app should be released for at least three years.
    \item[IC1.2] The release notes of this app should have less than 80\% (suggested) repetition rate after splitting into sentence levels.
    \item[IC1.3] The app should have a sufficient amount of release notes and user reviews.
\end{itemize}

\subsubsection{Data Processing}
\label{dp}

Our research dataset consists of app release notes and reviews of the same apps. Generally, the number of user reviews is much greater than that of the release notes. Meanwhile, app release notes are often written in a more regular and structured way. Therefore, our data processing is divided into two parts: user review processing and release note processing.

\textbf{Task 1: Data Processing for User Reviews} is conducted by the following three steps.

\begin{figure}[htbp]
  \includegraphics[scale=.6]{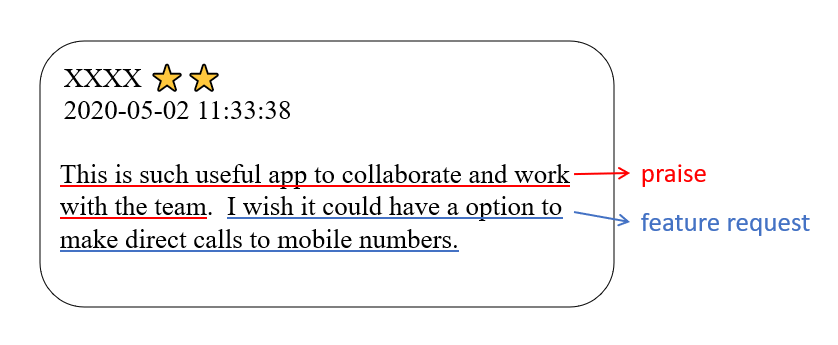}\\
  \centering
  \caption{Example of a user review that contains multiple pieces of information.}
  \label{fig:multiple}
\end{figure}

\textbf{\textit{Step 1.1: Sentence Splitting.} }One user review usually describes user feedback from  multiple aspects with several sentences. Taking the user review in Figure \ref{fig:multiple} as an example, it describes two types of feedback: the first sentence praises the app, while the second sentence requests a new feature.
In order to simplify the analysis of user reviews,  NLTK~\cite{NLTK} is used to split the collected user reviews into sentences to ensure each unit of user reviews only contains one piece of information.

\textbf{\textit{Step 1.2: Sentence pre-processing.}} In this step, multiple NLP (Natural Language Processing) techniques were applied to the textual content of app review sentences. Specifically, NLTK \cite{NLTK} was adopted to perform stopword removal, punctuation removal, and lemmatization.

\textbf{\textit{Step 1.3: Non-informative review sentences filtering.}} First, this paper follows the definition of informative reviews and non-informative reviews in ~\cite{ARMiner}: \textit{`informative'} implies that the review is constructive/helpful to app developers, and ``\textit{non-informative}'' means that the review contains no information that is useful for improving apps.
To remove non-informative reviews, this paper reused EMNB~\cite{EMNB}, i.e., the semi-supervised machine learning algorithm adopted in ARMiner~\cite{ARMiner}, as the classifier of user reviews.
 
Compared with supervised algorithms, EMNB fits the case of a small amount of manually-labeled data along with plenty of unlabeled data to achieve better performance. 
Specifically, we leverage the dataset provided by Chen et al.~\cite{ARMiner} as the training data to build the classifier for data filtering.

\textbf{Task 2: Data Processing for Release Notes}, which is conducted by the following two steps.

\textbf{\textit{Step 2.1: Sentence splitting.}} Similar to user reviews, a release note usually contains of several release sentences in a list format. As shown in Figure~\ref{fig:1.1} (a), each piece of information basically starts a new line with a special symbol (e.g., ``$-$'', ``$*$'', or ``$\cdot$''). We observed that some release notes do not end with a period at the end of each line, which may leads to failure in sentence-level splitting using the natural language processing toolkits, like NLTK or SpaCy~\cite{spaCy}, so we divide the release notes by paragraphs.

\textbf{\textit{Step 2.2: Duplication removal.}} 
This step can be used to clean duplicated release note sentences from the following three aspects.
First, as mentioned earlier, many apps tend to make `perfunctory' updates in a pattern that does not deliver the content of the updating, for example, ``\textit{Fixed stability issues (Sporify, 2021-11-11)}'', ``\textit{Bug fixes and performance improvements (Google Drive, 2020-10-27)}'', and ``\textit{Some bug fixes. (SHEIN, 2017-11-03)}''. These release note sentences occur frequently but seldom bring any details. Therefore, the removal of this type of release note sentences can greatly increase the quality of release notes to be included in the research dataset. Second, some release notes start or end with a welcome or appreciation statement at the beginning or the end of each update, or for example, ``\textit{Take collaboration to the next level by connecting over video with Google Meet, part of Google Workspace. (Google Meet, 2021-08-31)}'', and some release notes also tend to have a general statement before the list of new features, such as ``\textit{Some recent additions include: (Pandora, 2019-01-31)}'' and  ``\textit{This release brings enhancements that help you get more out of your inbox: (Gmail, 2018-08-13)}''. These release notes are also often duplicated, and the process can significantly reduce their weight in the update log sentences.. Third, release notes often contain major updates and subsequent minor updates, and a small number of apps repeat the content of major updates in minor updates for the purpose of reminding users of their new major features. The de-duplication of keeping only the initial occurrence of the release notes can ensure the removal of such duplicate information.

\subsubsection{Criteria of Release Notes}
\label{rnsc}

Due to the aforementioned characteristics of release notes, not all the release notes of selected apps meet the objective of this work. To construct the research dataset of this paper, the following two inclusion criteria are defined to support the second round selection applied in the release notes of the apps, which are the results of implementing the five ICs defined in Section~\ref{as}.

\begin{itemize}

    \item[IC2.1] The content of the sentence-level release note is understandable.
    \item[IC2.2] The sentence-level release note contains valuable updating information.

\end{itemize}

\subsection{RoseMatcher}
\label{rose}
This section provides the details of \textsc{RoseMatcher}, including text representation, sentence similarity calculation, and high-confidence matched pairs identification.

\subsubsection{Text Representation with Word Embeddings}
\label{trwe}
As the prerequisite of comparing the similarity between app reviews and release notes, their textual content should be transferred into numerical presentations. 
Word2Vec~\cite{Word2Vec} and BERT~\cite{devlin2019bert} are two landmark models in natural language processing. Both Word2Vec and BERT consider the context of certain word-for-word embeddings. Differently, BERT, based on Transformers~\cite{Transformer}, considers both the context (bidirectional encoder) and word order information to present the embedding of the word, while Word2vec does not consider the word order, and the relevance of all words in the whole sentence is not considered due to the limitation of window size. Moreover, the word vector represented by Word2Vec is static, while BEET represents the dynamic word vector derived from the context of the sentence. According to our requirements, the improvement of BERT compared with Word2Vec is relatively limited for the following reasons: (1) the research data are homogeneous in nature, which are all users' or developers' statements of app-related functions, etc, which leads to the low possibility of polysemy, and the improvement of dynamic representations is not significant. (2) it is the nature of BERT to consider the context of the sentence that leads to its ability to identify the difference of opposite meanings well (e.g., synonyms and antonyms and positive and negative sentiments), but user feedback towards app tend to have both positive and negative statements, so that the improvement of BERT compared to Word2Vec may conversely become a factor of lower matching rate. Therefore, we use both models in parallel for matching.

\textbf{Text Representation with SBERT}
For the SBERT model, we get the representation of sentence embedding using pre-trained Sentence BERT model \textit{all-MiniLM-L6-v2}, which is available on Huggingface~\cite{huggingface}. It uses the pre-trained model~\cite{wang2020minilm} and fine-tuned in on a 1,170,060,424 sentence pairs dataset and maps sentences \& paragraphs to a 384-dimensional dense vector space.

\textbf{Text Representation with Word2Vec}
For the Word2Vec model, we use 1,082,192 informative original user review sentences collected and processed by ourselves as training data, with the training mode of Skip-gram, to map each word to a 300 dimensional vector space. In our approach, we only include embeddings of nouns, verbs, and adjectives, which we can automatically detect with a part-of-speech (POS) tagger. Some previous studies have either retained only nouns~\cite{MatchBugReportsWithReviews} or only verbs and nouns~\cite{SupportingFeatures} when performing similar matching approaches, however, our observation shows that, for example, in the user reviews and release notes concerning \textit{dark mode}, if we remove the adjective \textit{dark} in the sentence \textit{We need dark mode support}, the matching of key information cannot be completed, similarly, the word \textit{left} in the sentence \textit{The bar should be placed to the left side} also plays a key role that cannot simply be removed.

Hence, we defined Formula (1) to represent sentence embeddings ($E_{sentence}$) using the Word2Vec model. 

\setlength{\mathindent}{0cm}
\begin{align}
E_{sentence}\!=\!w_{verb}\dfrac{{\sum}E_{verb}}{n_{verb}}\!+\!w_{noun}\dfrac{{\sum}E_{noun}}{n_{noun}}\!+\!w_{adj}\dfrac{{\sum}E_{adj}}{n_{adj}}
\end{align}

where \textit{$w_{verb}$}, \textit{$w_{noun}$} and \textit{$w_{adj}$} are weights of verbs, nouns and adjectives respectively, and $w_{verb}+w_{noun}+w_{adj} = 1$; \textit{$E_{POS}$} denotes the word embedding of a word whose part of speech is \textit{POS}; \textit{$n_{POS}$} denotes the number of words whose part of speech is \textit{POS}. Here, we assign $w_{verb}=w_{noun}=w_{adj} = \frac{1}{3}$.

\subsubsection{Sentence Similarity}
\label{sm}
Given the numerical representation of release note sentences and user review sentences, \textsc{RoseMatcher} utilizes sentence similarity to measure the relevance between them. Common methods of calculating sentence similarities, such as edit distance~\cite{Edit}, lexical overlap~\cite{2005Recognizing}, and substring matching~\cite{StringMatching}, can only work in some simple cases~\cite{kenter2015short}, and can not capture the semantic similarity between texts. Based on the word embedding representation, cosine similarity performs well in capturing semantic features, as a good word embedding model maps similar words to proximate vectors. Hence, the cosine similarity algorithm, as defined in Formula (2), is used to calculate the similarity between release note sentences and user review sentences. 

\setlength{\mathindent}{0.5cm}
\begin{align}
sim(E^{rn},E^{ur}) = \dfrac{\sum^{n}_{i=1}E^{rn}_iE^{ur}_i}{\sqrt{\sum^{n}_{i=1}(E^{rn}_i)^2}\cdot \sqrt{\sum^{n}_{i=1}(E^{ur}_i)^2} }
\end{align}

where $E^{rn}$ and $E^{ur}$ denote the sentence embedding representation of the given release note and user review respectively, and $n$ denotes the dimension of the sentence vectors.

\subsubsection{Identifying high-confidence matched pairs}
\label{ihcmp}
Given any release note sentence, we can calculate its similarity with the user review sentences of the app, and according to the ranking of similarity, we can obtain the top N matched pairs with the highest similarity. However, high similarity is not equivalent to high relevance. Our subsequent results also show that less than 1/3 of the matched pairs with high similarity are relevant ones when using a single model for matching. In order to efficiently filter out the truly relevant matched pairs, we creatively propose the method of intersecting of the top results obtained by two models to get high-confidence matched pairs, as shown in Formula (3). Although this method may drop some relevant matched pairs, it can greatly improve the efficiency of manual labeling to obtain more relevant matched pairs, which is of great value to our empirical study. 

\begin{equation}
    \Phi = \big\{p \vert p \in \text{topN}(\mathbf{P_1}) \big\} \cap \big\{ p \vert p\in \text{topN}(\mathbf{P_2}) \big\}
\end{equation}

where \textbf{$P_1$} and \textbf{$P_2$} denote the release note and user review pairs of two models respectively, topN() is the function to get the set of the top N pairs with the highest similarity, and $\Phi$ denotes the set of high-confidence matched pairs.

\subsection{Analyzing Matched Pairs}
\label{ra}
This section focuses on how we proceed with and from what perspective we analyze the high-confidence matched pairs that \textsc{RoseMatcher} outputs. We first manually label the matched pairs suggested by \textsc{RoseMatcher} to determine the truly relevant matched pairs and evaluate the effectiveness and accuracy of \textsc{RoseMatcher}. Subsequently, we identify the roles that user reviews play according to the relevance between release notes and user reviews based on the relevant matched pairs. Finally, we analyze the relevant matched pairs from the perspective of the release time of user reviews and release notes. These steps are described in detail at the methodological level as follows.

\subsubsection{Manual Labelling}
\label{ml}
Even though \textsc{RoseMatcher} generates high-confidence matched pairs, it is also necessary to label these pairs manually to accurately filter out the truly relevant matched pairs is indispensable.

In this paper, we propose the criteria to identify whether a user review is `relevant' to a certain release note, as outlined below. 

\begin{enumerate}
    \item The review explicitly refers to or talks about a specific feature, enhancement, or issue introduced in the corresponding release note.
    \item The review talks about the positive or negative feedback on the changes whose implementation in the update is mentioned in the release note, reflecting the user's experience or bringing constructive suggestions.
\end{enumerate}

Concretely, to establish these criteria for labeling relevant matched pairs, the authors conducted a thorough analysis of 300 randomly selected samples. A matched pair is considered \textit{relevant} if the content of the user review aligns with the release note. This can occur when the review directly corresponds to the release note content, suggesting that the review could be a driving factor for the release note, or when the review provides feedback based on requests or concerns stemming from the new update. On the other hand, a matched pair is labeled as \textit{irrelevant} if no discernible connection exists between the release note and the user review.

After manual labeling, we evaluate the effectiveness and accuracy of our approach. For this purpose, we define Hit Ratio, as described in Formula (4). We leverage the percentage of matched pairs labeled as `relevant' in the top 80 pairs as the measurement standard.

\setlength{\mathindent}{2.3cm}
\begin{align}
    Hit Ratio = \dfrac{|\mathbf{P_r}|}{|\mathbf{P_r}|+|\mathbf{P_{ir}}|}
\end{align}

where \textbf{$P_{r}$} and \textbf{$P_{ir}$} are the set of relevant and irrelevant matched pairs of a certain app or some apps respectively.

\subsubsection{Identifying the Roles of User Reviews}
\label{rur}
In order to investigate the role that user reviews play in app updates from the perspective of release notes, we analyze the content and the relevance of all the matched pairs which are manually marked as ``relevant'' in Section ~\ref{ml} with an iterative technique. First, we draw a stratified random sample~\cite{SelectingEmpiricalMethods} of the relevant matched pairs. Stratified sampling can divide the data into manually exclusive groups according to the apps. Next, the first and the third authors independently label the random sample to identify the roles user review play in app update. Multiple roles are not allowed since we split both release notes and user reviews into sentences, and a sentence usually takes on only one role.

The labeling process includes two iterations on one single matched pair. To start with, We create an empty set of role labels. Next, we carefully read the contents of both the release note and user review in each matched pair. When a certain type of role that the user review plays in this app update is identified, this role will be added to the set of role labels. We repeated this operation in each of the matched pairs till no new role could be added to the role set. 

In this way, the initial list of role types is created and then shared among the first three authors to include new types and dispute duplicated types.

Based on stratified sampling, we can initially define the role types, and the first and third authors labeled all remaining matched pairs independently after reaching an agreement on the initially defined role types. During the labeling process, once a new type of matched pair is encountered, it will be immediately proposed and discussed by the first, second, and third authors whether the initially defined classification criteria need to be modified.

\subsubsection{Time Interval Calculation}
\label{tdrr}
For any matched pair, the lengths of time interval ($\Delta t$) between the posting time of a release note ($t_{rn}$) and that of its matched app reviews ($t_{ur}$) are easy to observe and collect, and the time interval between them can reflect, to some extent, the attention the development teams pay to user reviews and the attention users pay to release notes. Therefore, for each matched pair, we calculate the time interval between the release note and user review using Formula (5).
\setlength{\mathindent}{3.2cm}
\begin{equation}
    \Delta t = t_{rn} - t_{ur}
\end{equation}

where $t_{rn}$ and $t_{ur}$ denote the posting time of the release note and that of the review in one matched pair respectively, and $\Delta t$ denotes the time interval value (unit: days).

Note that in this paper, a negative $\Delta t$ means that in this matched pair, the relevant user review was posted before the release note. Whereas, the positive value of $\Delta t$ means that the relevant user review was posted after this release note. Furthermore, if we use \textbf{T} to represent the set of all the $\Delta t$, the following formulas (6) and (7) can be used to calculate the average time interval value (unit: days) between release notes and the reviews posted before and after the release notes, respectively. This indicates the degree of attention that the development teams pay to user reviews and the degree of attention users pay to release notes to some extent.

\setlength{\mathindent}{1.8cm}
\begin{align}
\overline{T}_{before} =\dfrac{ -\Sigma_{time\in{t\in \mathbf{T} \vert t<0}}time}{\vert {t\in \mathbf{T} \vert t<0}\vert}
\end{align}

\begin{align}
\overline{T}_{after} =\dfrac{ \Sigma_{time\in{t\in \mathbf{T} \vert t>0}}time}{\vert {t\in \mathbf{T} \vert t>0}\vert}
\end{align}

\section{Experiments and Results}
\label{er}
Our work aims to identify the impact of user reviews on app release notes, which is based on the identification and analysis of the one-to-one matched pairs of app reviews and release notes. For this purpose,  we designed the following three research questions (RQs):

\textbf{RQ1: }Can \textsc{RoseMatcher} serve to identify the one-to-one matched pairs of user reviews and release notes effectively and accurately?

\textbf{RQ2: }What roles do user reviews play in app updates according to the matched pairs?

\textbf{RQ3: }
What is the impact of user reviews on release notes from the perspective of their posting time?

\begin{table*}[t]
\centering

\label{tab_dataset}
\begin{tabular}{ll|lll|lll}
\hline
\multirow{3}{*}{\textbf{App Name}} & \multirow{3}{*}{\textbf{Category}} & \multicolumn{3}{c|}{\textbf{Release Notes}}                                                                                               & \multicolumn{3}{c}{\textbf{User Reviews}}                                                                                                     \\ \cline{3-8} 
                          &                           & \multirow{2}{*}{\textbf{No. of RN}} & \multicolumn{2}{c|}{\textbf{No. of RN Sentences}}        & \multirow{2}{*}{\textbf{No. of UR}} & \multicolumn{2}{c}{\textbf{No. of UR Sentences}}             \\ \cline{4-5} \cline{7-8} 
                          &                           &                                                      & \textbf{Total} & \textbf{De-duplicated} &                                                      & \textbf{Total}     & \textbf{Informative} \\ \hline
Reddit                    & News                      & 233                                                  & 450                             & 218                                     & 62,598                                               & 129,545                             & 41,110                                  \\
Spotify                   & Music                     & 182                                                  & 506                             & 50                                      & 368,243                                              & 808,441                             & 300,478                                 \\
Pandora                   & Music                     & 115                                                  & 235                             & 116                                     & 107,241                                              & 241,497                             & 79,906                                  \\
SHEIN                     & Shopping                  & 161                                                  & 361                             & 88                                      & 45,516                                               & 117,529                             & 21,271                                  \\
Instagram                 & Photo\&Video              & 253                                                  & 380                             & 77                                      & 461,264                                              & 918,729                             & 600,299                                 \\ \hline
\multicolumn{2}{c|}{\textit{Total}}                            & \textit{944}                      & \textit{1,805} & \textit{549}         & \textit{1,046,862}                  & \textit{2,035,741} & \textit{1,043,591}     \\ \hline
\end{tabular}
\caption{Summary of the collected and processed app data. Release Notes (RN); User Reviews (UR).}
\end{table*}

\subsection{Dataset}
\label{dc}
The collection of our dataset strictly follows the app selection criteria proposed in Section~\ref{as}. We glanced at Top 2 apps of each of the 26 categories in the Apple App Store~\cite{AppStore} except the `Game' category, and finally filtered out 5 suitable apps. The reason to exclude Game apps is that although the `Game' category has the highest percentage of apps in the App Store (21.07\%)~\cite{AppPercentage}, many games provide their own forums or social media platforms to release new versions, rather than in app repositories. Additionally, the data sourced from the apps in the `Game' repository usually contain game terminologies in plenty, which may require researchers to spend much time experiencing the game to understand the rules, terminology, etc. in order to read the release notes and user reviews.


Finally, app reviews and release notes of five apps, i.e., \textit{Reddit}, \textit{Spotify}, \textit{Pandora}, \textit{SHEIN} and \textit{Instagram} in Apple App Store, are included to construct the dataset of this paper. Data collection was performed in February 2022, and we crawled all the user reviews and release notes of these 5 apps posted from January 1st, 2017 to January 1st, 2022. The raw dataset of this study has been made available online~\cite{dataset}. Subsequently, we process the collected release notes and user reviews respectively according to the steps described in Section~\ref{dp}. Table~\ref{tab_dataset} summarizes the size of the collected and processed app data of these five apps.

\subsection{Answer to RQ1}
\label{arq1}

\begin{table*}[htbp]
\centering
\label{tab_rq1}
\begin{tabular}{l|lll|lll|lll} 
\hline
\multirow{2}{*}{App Name} & \multicolumn{3}{c|}{Word2Vec}                  & \multicolumn{3}{c|}{SBERT}                     & \multicolumn{3}{c}{Word2Vec $\cap$ SBERT}            \\ 
\cline{2-10}
                          & \#             & Hit Num      & Hit Ratio      & \#             & Hit Num      & Hit Ratio      & \#           & Hit Num      & Hit Ratio       \\ 
\hline
Reddit                    & 240            & 64           & 0.267          & 240            & 79           & 0.329          & 47           & 29           & 0.617           \\
Spotify                   & 240            & 120          & 0.500          & 240            & 123          & 0.513          & 69           & 53           & 0.768           \\
Pandora                   & 240            & 42           & 0.175          & 240            & 44           & 0.183          & 51           & 17           & 0.333           \\
SHEIN                     & 240            & 4            & 0.017          & 240            & 8            & 0.033          & 80           & 4            & 0.050           \\
Instagram                 & 240            & 186          & 0.775          & 240            & 171          & 0.713          & 54           & 52           & 0.963           \\ 
\hline
\textit{Total}            & \textit{1,200} & \textit{416} & $\phi$\textit{0.347} & \textit{1,200} & \textit{425} & $\phi$\textit{0,354} & \textit{301} & \textit{155} & $\phi$\textit{0.515}  \\
\hline
\end{tabular}
\caption{Performance (hit ratio) of the three models (single model using Word2Vec, single model using SBERT and proposed model combining Word2Vec and SBERT). }
\end{table*}

\subsubsection{Research Data for RQ1}

As mentioned in Section ~\ref{dc}, our dataset covers all user reviews and release notes of the five selected apps from 2017 to 2021. Then, for each of these five apps, we randomly selected one release note in 2018, 2019, and 2020 respectively for subsequent matching. This brings a set of 15 (3 years $\times$ 5 apps) sentence-level release notes, in order to ensure that any app release note will have at least one year of reviews before and after its release time respectively. For each release note sentence, we set \textit{N} in Section \ref{rur} as 80 to choose the Top 80 matched pairs generated by \textsc{RoseMatcher}. Since \textsc{RoseMatcher} employed two models, i.e., the final dataset serves to answer RQ1 is 2400 selected matched pairs (15 release notes $\times$ 80 matched pairs $\times$ 2 models), and these matched pairs will be analyzed manually.

Our selection strategies for this study, e.g., a random selection of release notes and set N as 80 in Section~\ref{rur}, were proposed to balance several factors. First, we aimed to guarantee a sufficient number of reviews both before and after each selected release note, in order to thoroughly analyze the relation between user reviews and release notes. Second, we needed to manage the workload of manually labeling matched pairs. To address these requirements effectively, we employed a sampling strategy where we randomly chose one release note per year for each app across a three-year span and set N equal to 80, which enabled us to acquire a manageable number of matched pairs (2,400 in total), while maintaining adequate coverage of the release notes from all selected apps.

\subsubsection{Results for RQ1}
\begin{figure}[htbp]
  \includegraphics[scale=.5]{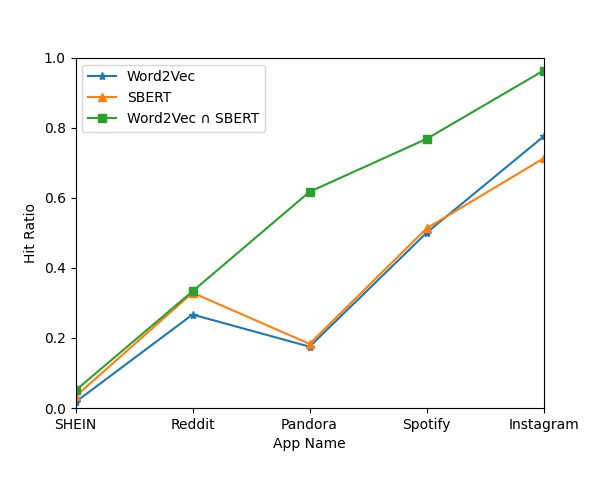}\\
  \centering
  \caption{Hit ratio of leveraging Word2Vec single model, SBERT single model, and the combination of Word2Vec and SBERT models. }
  \label{fig:hit}
\end{figure}

\begin{figure}[htbp]
  \includegraphics[scale=.5]{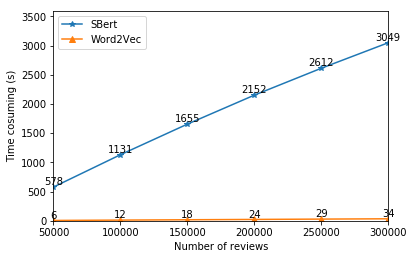}\\
  \centering
  \caption{Time (unit: seconds) consumed for text encoding and similarity calculation (Word2Vec vs SBERT).}
  \label{fig:time}
\end{figure}

Table~\ref{tab_rq1} summarizes the overall results of the evaluation. In the table, we compare the performance of two models, Word2Vec and SBERT, in matching relevant pairs, and we also give the hit ratio of \textsc{RoseMatcher} by taking the intersection of high-similarity matched pairs output by Word2Vec and SBERT. Figure~\ref{fig:hit} visualizes the hit ratio of leveraging two single models and two models simultaneously by taking the intersection. Figure~\ref{fig:time} shows the time that the Word2Vec model and SBERT model consume for encoding and similarity calculation.

Our results show that although the large model (SBERT) trained on a large dataset performs slightly better than the small self-trained model (Word2Vec), the SBERT model consumes much more time (nearly 10 time) for encoding and computing similarity than the Word2Vec model. In addition, the hit ratios of the two single models are both around 0.35, while combining two models by taking the intersection can achieve the hit ratio of 0.515, although this approach may discard some matched pairs that should be labeled as relevant, this can greatly improve the efficiency of manual labeling and thus expand the data set to compensate for the discarded samples.

\begin{center}
\begin{tcolorbox}[colback=gray!10,
                  colframe=black,
                  width=8cm,
                  arc=1mm, auto outer arc,
                  boxrule=0.5pt,
                 ]
\textbf{Key Findings: }
In this section, we manually label a total of 2400 high similarity matched pairs matched by 18 release note sentences, and the results show that, overall, SBERT (0.354) has a slightly higher hit ratio  than Word2Vec (0.347) if a single model is used, but SBERT consumes nearly 10 times more time, while taking intersection to combine Word2Vec and SBERT greatly outperforms SBERT and Word2Vec single model, which can reach a hit ratio of 0.515.
\end{tcolorbox}
\end{center}

\subsection{Answer to RQ2}
\label{arq2}

\subsubsection{Research Data for RQ2}
The results of RQ1 show that high-confidence matched pairs can be obtained by combining Word2Vec and SBERT to take the intersection, which constitutes \textsc{RoseMatcher}. This discards matched pairs that are only matched by a single model but give us the ability to expand the investigated release note entries.

For this purpose, we reselect the investigated release note sentences according to the release note selection criteria described in Section~\ref{rnsc}. Specifically, we randomly select one release note in each quarter of 2018, 2019, and 2020 respectively as the investigated release note sentences, i.e., a total of 60 (4 quarters $\times$ 3 years $\times$ 5 apps) release notes. If a release note sentence that meets the selection criteria cannot be picked up within a certain quarter, we will randomly select one within its adjacent quarter to try to ensure an even distribution of the 12 release note sentences collected for each app. 

We take 60 release note sentences and their corresponding sets of user review sentences as input to \textsc{RoseMatcher}, and for each release note, we select the top 80 matched pairs with the highest similarity. Eventually, \textsc{RoseMatcher} outputs a total of 984 high-confidence matched pairs. Similar to the evaluation experiment, the authors labeled the matched pairs according to the labeling criteria and methods described in Section~\ref{ml}. Table~\ref{tab_rq2} summarizes the Number of \textsc{RoseMatcher} outputs, hit numbers, and hit ratios of each app.

\subsubsection{Results for RQ2}

After manual labeling, our results show that the average hit ratio of our proposed \textsc{RoseMatcher} can reach 0.718, after we fourfold the number of investigated release note sentences. The lowest hit ratio is 0.545 and the highest reaches 0.901. Of the 984 suggested high-confidence matched pairs, 707 are labeled as relevant matched pairs. Subsequently, we identify the roles that user reviews play in app updates based on the 707 matched pairs, according to the methodology introduced in Section~\ref{rur}. Table~\ref{tab_examples} shows the 8 roles and gives some examples and the frequency for each role.

\begin{table}[h]
\centering
\label{tab_rq2}
\begin{tabular}{llll} 
\hline
App Name       & No. of Output  & Hit Num      & Hit Ratio       \\ 
\hline
Reddit         & 222            & 121          & 0.545           \\
Spotify        & 191            & 157          & 0.822           \\
Pandora        & 206            & 172          & 0.835           \\
SHEIN          & 223            & 129          & 0.578           \\
Instagram      & 142            & 128          & 0.901           \\ 
\hline
\textit{Total} & \textit{984} & \textit{707} & $\phi$\textit{0.718}  \\
\hline
\end{tabular}
\caption{Summary of the number of matched pairs and hits for the data of RQ2.}
\end{table}

\begin{table*}[!ht]
\label{tab_examples}
{\footnotesize
\begin{tabular}{lm{1.5cm}m{1.1cm}m{4.3cm}m{6.8cm}l} 
\hline
\textbf{\#} & \textbf{Role} & \textbf{App} & \textbf{Release Note} & \textbf{User Review} & \textbf{\%} \\ 
\hline
\multirow{7}{*}{1}  & \multirow{7}{1.5cm}{Feature Requester}            & Reddit                     & Mods can now edit post flairs in their communities                                                               & I'd like to be able to edit my subreddit flair.                                                                                                        & \multirow{7}{*}{26.7\%}  \\
                    &                                             & \multirow{3}{*}{Spotify}   & \multirow{3}{4.2cm}{We've now added support for the iPhone X}                                                        & But please add support for the iPhone X.                                                                                                               &                       \\
                    &                                             &                            &                                                                                                                  & Please do the right thing and support the screen size for the iPhone X.                                                                                &                       \\ &  & SHEIN & Added a translate function for comments, so now you can read every comment & Also I wish the comments can be translated \\
\hline
\multirow{5}{*}{2}  & \multirow{5}{1.5cm}{Bug Reporter}                 & Pandora                    & Now you can ask Siri to play your music on Pandora                                                               & You say Hey Siri resume Pandora and that idiot will play some random station just on a whim.                                                           & \multirow{5}{*}{20.2\%}  \\
                    &                                             & \multirow{4}{*}{Spotify}   & \multirow{4}{4.2cm}{Fixed issue when clicking external links}                                                        & When i try to click the link, it says cant open link on this device.                                                                                   &                       \\
                    &                                             &                            &                                                                                                                  & Please fix this because it's really frustrating to not be able to open any Spotify links.                                                              &                       \\ 
\hline
\multirow{3}{*}{3}                  & \multirow{3}{1.5cm}{Complainer}                                   & SHEIN                      & Try out the search function on new product page, find products you like more easily                              & It's very annoying because I have to search all over creation to find the items again                                                                  & \multirow{3}{*}{20.2\%}                   \\ &  & Pandora & Introducing support for iPhone X  & I refuse to use this app until iPhone X support is added \\
\hline
\multirow{3}{*}{4}  & \multirow{3}{1.5cm}{Praiser}                     & Spotify                    & Low data mode on iOS Spotify will now turn on Data Saver when your device s Low Data Mode is turned on           & The main thing for me is that Spotify has a data saver to help lower the data used when streaming                                                      & \multirow{3}{*}{11.9\%}   \\
                    &                                             & Pandora   & Go to Settings and choose the new Change App Icon option                                       & I also love the fact that you can change the app icon                                                                                                  &                       \\
\hline

\multirow{6}{*}{5}  & \multirow{6}{1.5cm}{Quality Issue Raiser}             & Spotify                    & Try saying Hey Siri Play music on Spotify or any other voice query of your choice adding on Spotify at the end   & Also asking Siri to play a song on Spotify sometimes doesn't work                                                                                      & \multirow{6}{*}{11.4\%}   \\
                    &                                             & Pandora                    & Introducing our new standalone Apple Watch app                                                                   & Come on where is the Apple Watch app                                                                                                                   &                       \\
                    &                                             & Instagram                  & Introducing new Messenger features in Instagram                                                                  & Indeed your update is just a lie because there's no new messenger features on Instagram                                                                &                       \\ 
\hline

\multirow{4}{*}{6}  & \multirow{4}{1.5cm}{Dispraiser}                  & Instagram                  & New Settings Menu We ve made it easier to find your settings like notifications and account controls             & i don't like the new way to go to settings                                                                                                             & \multirow{4}{*}{4.0\%}   \\
                    &                                             & Pandora                    & Go to Settings and choose the new Change App Icon option                                                         & Honestly I don't need the fancy change your icon updates                                                                                               &                       \\ 
\hline

\multirow{8}{*}{7}  & \multirow{8}{1.5cm}{Subsequent Feature Requester} & Reddit                     & Drafts are now available                                                                                         & But one thing I would like to see improved is for a way to draft picture and video posts instead of discarding it completely                           & \multirow{8}{*}{3.8\%}   \\
                    &                                             & \multirow{2}{*}{Instagram} & \multirow{2}{*}{Introducing Voice Messaging}                                                                     & But what about commenting a voice message                                                                                                              &                       \\
                    &                                             &                            &                                                                                                                  & you should be able to listen to the voice message before you send it                                                                                   &                       \\
                    &                                             & Spotify                    & Try saying Hey Siri Play music on Spotify or any other voice query of your choice adding on Spotify at the end   & Apple could allow better voice controls for Spotify it would be nice to be able to add songs to playlists or take songs off of playlist by asking Siri &                       \\
\hline

\multirow{3}{*}{8}  & \multirow{3}{*}{Questioner}                   & \multirow{3}{*}{Instagram} & \multirow{2}{4.2cm}{Instagram now filters out bullying comments intended to harass or upset people in our community} & I feel like restricting bullying comments will just cause more problems within the instagram community                                                 & \multirow{3}{*}{1.7\%}    \\
                    &                                             &                            &                                                                                                                  & Now filters out bullying comments intended to harass or upset people. How true is this?                                                                &                       \\ 
\hline

\end{tabular}}
\caption{Examples of 8 user review roles and the corresponding percentages.}
\end{table*}

\textbf{Role 1:} \textsc{\textbf{Feature Requester}} (189 matched pairs, 26.7\%)

User reviews as feature requester are often written in a requesting tone, expressing the users' request for certain features that the app does not equip with. From the perspective of release notes in the matched pairs, release notes addressed the requesting user proposed in the user reviews. Role 1 is the most common role played by user reviews. See one example in Table~\ref{tab_examples}, ``\textit{But please add support for the iPhone X}'' and ``\textit{Also I wish the comments can be translated}'' are the feature requests for \textit{iPhone X} and ``\textit{Comment Translation}'' respectively, which are all introduced in the subsequent release notes. Note that we cannot assert whether these user reviews directly led the development team to make the relevant update, these user reviews can be deemed as one of the important potential evolutionary requirements of the updates introducing new features.

\textbf{Role 2: }\textsc{\textbf{Bug Reporter}} (143 matched pairs, 20.2\%)

User reviews as bug reporter articulate the problems that users encounter when using the apps, outlining information about what is wrong and needs fixing. From the perspective of app release notes, the app development teams fix the bug reported by users. For example, the user review -- ``\textit{When i try to click the link, it says can't open link on this device}'' reports a bug that was fixed in the release note -- ``\textit{Fixed issue when clicking external links}''. Similar to user reviews as feature requester, the bugs or terrible experiences raised from these user reviews prospectively can be detected by the development teams, thus helping smooth out software. 

\textbf{Role 3: }\textsc{\textbf{Complainer}} (143 matched pairs, 20.2\%)

User reviews as complainer contain strong negative emotions, complaining about bad experiences when using the app. We tend to view these user reviews as indirect bug reports or feature requests since negative emotions often source from the terrible experience of using the software, such as annoying bugs and the absence of desired features, and developers tend to address and respond to these specific complaints. See the example in Table~\ref{tab_examples}, ``\textit{It’s very annoying because I have to search all over creation to find the item}'' complains about the difficulty in finding items, which is solved by introducing the new ``\textit{search function}''. Note that such user reviews can often be sideways feature requester or bug reporter, but as long as the review is a complaint with negative sentiment, we consider it a complainer.

\textbf{Role 4: }\textsc{\textbf{Praiser}} (84 matched pairs, 11.9\%)

User reviews as praiser give positive feedback according to the content of the release notes, expressing love for the app new features or appreciation for bug fixes, etc. From the perspective of app release notes, user reviews as praiser are positive responses from users to app release notes. See the examples in Table~\ref{tab_examples}, ``\textit{I also love the fact that you can change the app icon}'' directly expressed love for the new features -- ``\textit{Change App Icon}'', which is introduced in the release note. Such user reviews reflect the level of user satisfaction with app updates, and development teams can learn from these successful updates.

\textbf{Role 5: }\textsc{\textbf{Quality Issue Raiser}} (81 matched pairs, 11.4\%)

User reviews as Quality Issue Raiser encompass many categories, and these reviews are usually related to feedback from users about the quality of the update. Unlike praise and dispraise, they often do not simply express affection/commendation or discontent/censure on the updates, but rather propose problems that the updates bring. From the perspective of app release notes, user reviews as quality issue raiser are neutral responses from users to app release notes. For example, the release note introduces the ``\textit{Siri integration}'', and the relevant user review ``\textit{Also asking Siri to play a song on Spotify sometimes doesn’t work}'' conveys quality feedback from users on the usability aspects of this new feature. These reviews often contain more valid information because their feedback is more specific and the development team can make subsequent updates to improve new features based on them. For example, if a user complains that he/she cannot find the updated new feature, the development teams can consider adding instructions or making the new feature more visible in subsequent updates to improve user experiences

\textbf{Role 6:} \textsc{\textbf{Dispraiser}} (28 matched pairs, 4.0\%)

User reviews as dispraiser are criticism, censure or dislike of the content of the release notes, expressing dissatisfaction with new updates. From the perspective of app release notes, user reviews as dispraiser are negative responses from users to app release notes. See some examples in Table~\ref{tab_examples}, ``\textit{Honestly I don’t need the fancy change your icon update}'' conveys that the user considers the new feature of this update - ``\textit{Change App Icon}'' - to be meaningless, expressing his\/her displeasure with the update. Such reviews can reflect users' attitudes toward the updates, and development teams can then reduce their input on similar updates rendering users feel discontented.

\textbf{Role 7:} \textsc{\textbf{Subsequent Feature Requester}} (27 matched pairs, 3.8\%)

User reviews as subsequent feature requester are often constructive suggestions or creative ideas based on the change of new updates. In contrast, these user reviews are feature requests essentially but are distinguished from the feature requester defined in Role 1, since from the perspective of app release notes, feature requests are requirements raised by users that are addressed in subsequent release notes, while subsequent feature requests are constructive advice proposed by users based on the update. See the examples in Table~\ref{tab_examples}, after the introduction of ``\textit{Voice Message}'', a user came up with the idea to ``\textit{comment a voice message}'' and another user requested that ``\textit{they should be able to listen to the voice message before sending}''.

\textbf{Role 8:} \textsc{\textbf{Questioner}} (12 matched pairs, 1.7\%)

User reviews as questioner are doubts and questions from users towards app release notes, which mainly contains two main aspects: one is questioning of the effectiveness and the other is questioning of the consequence. Take the following release note as an example: ``\textit{Instagram now filters out bullying comments intended to harass or upset people in our community}''. On the one hand, ``\textit{Now filters out bullying comments intended to harass or upset people. How true it is?}'' is a question about the effectiveness of the update, since the user does not believe that the new feature is feasible; On the other hand, ``\textit{I feel like restricting bullying comments will just cause more problems within the instagram community}'' questions about the consequence that brings by the update, since this new update seems to bring some new bad influence in the user's opinion. Role 8 is the least common role that user reviews play.

\begin{center}
\begin{tcolorbox}[colback=gray!10,
                  colframe=black,
                  width=8cm,
                  arc=1mm, auto outer arc,
                  boxrule=0.5pt,
                 ]
\textbf{Key Findings: }
In this section, we identify 8 roles of user reviews according to their relevance to app release notes. User reviews dominantly play the roles of feature requester (26.7\%), bug reporter (20.2\%), and complainer (20.2\%) in app updates, which can be deemed as potential evolutionary requirements for development teams to update apps and deliver the corresponding release notes to users. Part of user reviews give praise (12.9\%) or dispraise (4\%) on app updates, and some users also provide feedback on the quality (11.4\%) of the updated content or raise questions (1.7\%). Users also submit subsequent feature requests (3.8\%) based on the updated content, containing constructive suggestions or creative ideas.
\end{tcolorbox}
\end{center}

\subsection{Answer to RQ3}
\subsubsection{Research Data for RQ3}
We reused the 707 relevant matched pairs suggested in Section~\ref{arq2} as the research data for answering RQ3.

\subsubsection{Results for RQ3}
\begin{figure*}[h]
\centering
  \includegraphics[width=\textwidth]{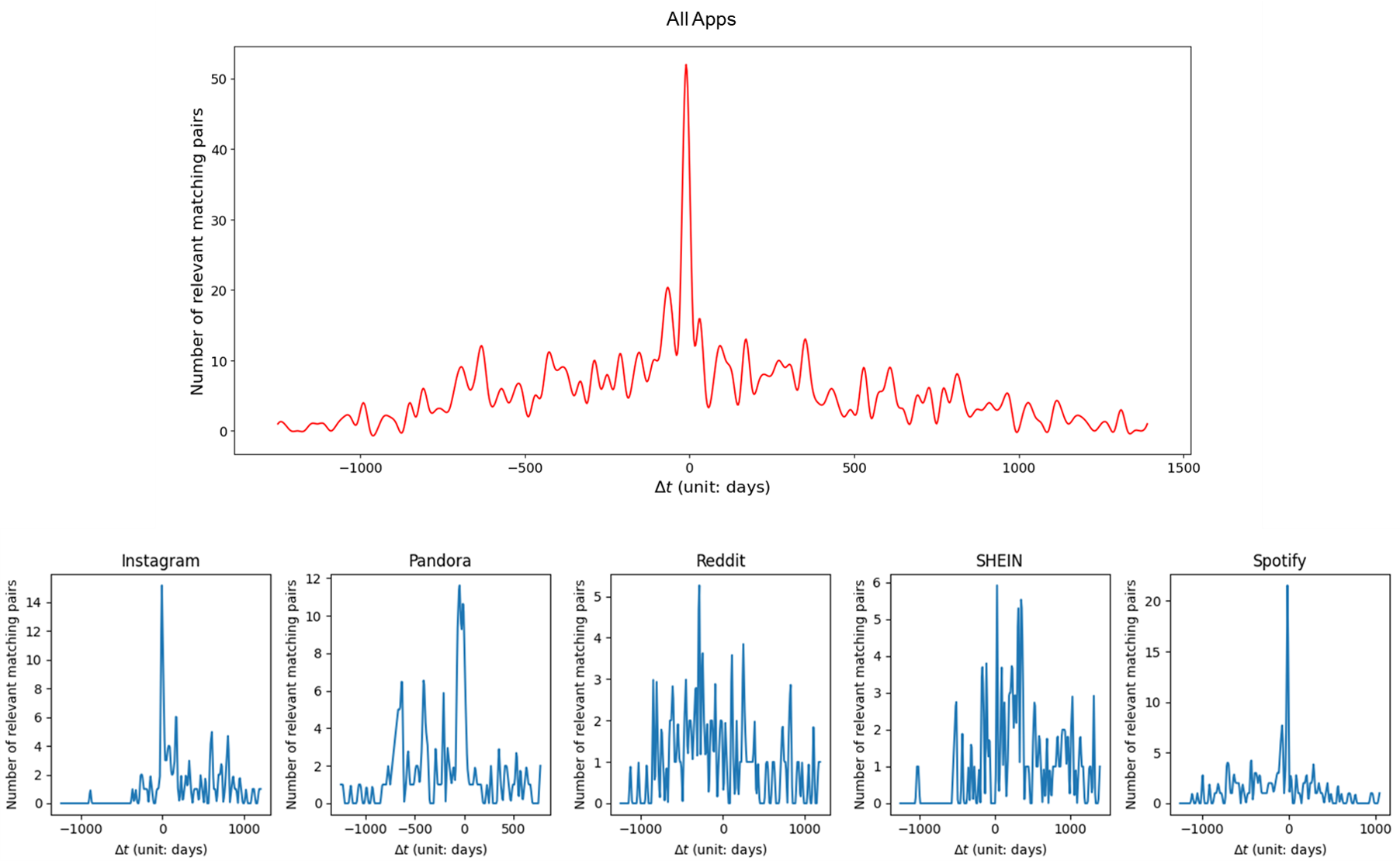}\\
  \caption{Time interval value distribution of all the relevant matched pairs (top figure with red line) and the relevant matched pairs for each app (bottom five figures with blue line), Instagram, Pandora, Reddit, SHEIN, and Spotify from left to right.}
  \label{fig:rq3_1}
\end{figure*}

As described in Section~\ref{tdrr}, we calculate the time interval value ($\Delta t$) for each matched pair of all the investigated apps (except ZOOM). Due to the dispersed $\Delta t$, we choose 20 days as an interval and visualized the number of data within each interval with $\Delta t$ as the x-axis and quantity as the y-axis for quantity statistics, as shown in the top figure with the red line in Figure~\ref{fig:rq3_1}.

Figure~\ref{fig:rq3_1} shows a clear concentration of $\Delta t$ between -80 and 0, suggesting that for release notes, relevant user reviews have a tendency to be concentrated within a certain time period prior to their release. Also, $\Delta t$ tends to be evenly distributed among the rest, but in general, the closer to the release notes, the relatively higher the number of relevant reviews is. Our results also show that the time interval value between a small number of relevant reviews and updated logs is even three to four years (maximum is 1,398 days).

The bottom five sub-figures with blue lines in Figure~\ref{fig:rq3_1} show the time interval value distribution of the relevant matched pairs for each app separately. The results show that \textit{Instagram}, \textit{Pandora} and \textit{Spotify} share similar characteristics, where the time interval value of their relevant matched pairs has a significant concentration around 0. Differently, \textit{Reddit} and \textit{SHEIN} have a more even distribution of the time interval value of their relevant matched pairs.

According to the formula (6, 7) mentioned in Section~\ref{tdrr}, for the matched pairs that user reviews posted before the corresponding release notes, the average time interval value ($\overline{T}_{before}$) is 365.59 days, while for the matched pairs that user reviews posted after the corresponding release notes, the average time interval value ($\overline{T}_{after}$) 430.33 days. 

\begin{center}
\begin{tcolorbox}[colback=gray!10,
                  colframe=black,
                  width=8cm,
                  arc=1mm, auto outer arc,
                  boxrule=0.5pt,
                 ]
\textbf{Key Findings: }
In this section, we visualize the time interval values between the release notes and user reviews in each relevant matched pair to explain the impact of user reviews on release notes on the dimension of their posting times. Our results show that there is a noticeable concentration of relevant reviews a short time before the release notes are posted. On average, for the matched pairs that user reviews posted before the release notes, the time interval value is about 1 year (366 days), while for the matched pairs that user reviews posted after the release notes, the time interval value is about 1 year and 2 months (430 days).
\end{tcolorbox}
\end{center}

\section{Interpretation of Results}
\label{discussion}

\subsection{Semantic Match Assessment between Release Notes and User Reviews}

\textbf{\textsc{RoseMatcher} effectively identifies semantic correspondences between release notes and user reviews.} Development teams and users often use different terminology to refer to the same features. However, by leveraging word embeddings, semantic matching can be achieved by calculating similarity. 
One of the strengths of \textsc{RoseMatcher} is its ability to detect well-matched pairs of phrases with identical meanings. Table~\ref{tab_semantic} provides examples of key phrases that express the same semantics but are expressed differently, which \textsc{RoseMatcher} is able to identify.

\begin{table}[!htbp]
    \centering
    \label{tab_semantic}
    \begin{tabular}{p{3.5cm}p{3.5cm}}
    \hline
    \textbf{Key phrase in release notes} & \textbf{Key phrase in user reviews}\\
    \hline
    voice & Siri \\
    Siri & voice command \\
    dark mode & night mode, dark display \\
    translate function & translator button \\
    freeze & get stuck, crush \\
    edit flairs & change flair \\
    \hline
    \end{tabular}
    \caption{Examples of the semantically similar key phrases detected by \textsc{RoseMatcher}.}
\end{table}

As a concrete example, if a release note announces that ``\textit{Our app now supports \textbf{Dark Mode} in iOS}'', \textsc{RoseMatcher} can successfully match user reviews that contain the exact key phrase ``Dark Mod'', such as ``\textit{needs a \textbf{dark mode}}'' and``\textit{I wish there was a \textbf{dark mode} feature}''. It can also identify user reviews that express the same idea in different words, such as ``\textit{You guys should add a \textbf{night mode} since the screen is too bright even when I switch it to Apple's \textbf{night mode}}'' and ``\textit{Also love the background displays for each album cover's artwork and with the new update for iPhone users when using the \textbf{dark display}}''. Another example would be a release note announcing ``\textit{Added a \textbf{translate function for comments} so now you can read every comment}''. \textsc{RoseMatcher} can match this with reviews such as ``\textit{Also, I wish the \textbf{comments can be translated}}'' and ``\textit{A translator button would be so helpful!}''.

Overall, \textsc{RoseMatcher} is able to achieve semantic matching to a significant extent, and the ability to detect different expressions of the same semantic concept is highly valuable.

\subsection{Possible Reasons for Irrelevant Matched Pairs}

\begin{table*}[h]
    \centering
    \label{tab_irrelevant}
    \begin{tabular}{llp{3.6cm}p{3.6cm}p{3.6cm}p{2.3cm}}
    \hline
    \textbf{\#} & \textbf{App} & \textbf{Release Note} & \textbf{Relevant Review} & \textbf{Irrelevant Review} & \textbf{Possible Reason} \\
    \hline
     1 & Spotify & Version of Spotify via \textbf{Siri }on \textbf{Apple Watch} (watchOS) is available for your enjoyment & Wish I could control Spotify with\textbf{ Siri} on \textbf{Apple Watch}  & Needs \textbf{Apple Watch }support to play music through Apple Watch and store music on \textbf{Apple Watch}  & Partial Matching \\
     \hline
     2 & Reddit & Disable \textbf{reply notifications} per individual \textbf{comment} or \textbf{post} & I still want to get \textbf{notifications} from \textbf{replies} of my \textbf{post} or \textbf{comments}  &  You can't \textbf{reply} to certain \textbf{comments}  & Partial Matching \\
     \hline
     3 & Instagram & Introducing new Messenger features in Instagram & Indeed your update is just a lie because there's no new messenger features on Instagram & So instagram got a new feature  & Partial Matching \\
     \hline
     4 & Spotify & We've now added \textbf{support} for the i\textbf{Phone X }&  \textbf{iPhone x }\textbf{support }took forever & Please update the app to\textbf{ support} the \textbf{iPad Pro} & Semantic Matching \\
     \hline
     5 & Reddit & Mods can now \textbf{edit post flairs} in their communities & I d like to be able to \textbf{edit my subreddit flair} & I like that they added the feature to \textbf{add flair to your posts} from the mobile app & Semantic Matching \\
     \hline
     6 & Instagram & We're introducing the ability to follow \textbf{hashtags} giving you new ways to discover photos videos and people on Instagram & N/A & And the new suggested photos after of the people you follow and give Instagram ads & Special Symbols or Numbers \\
     \hline
     7 & Spotify & Upcoming versions will only support \textbf{iOS 11} and later & N/A & Needs to be updated for full iPad support  & Special Symbols or Numbers \\
     \hline
     8 & SHEIN & More community activities added & N/A & Can you add create a list for me and for other & Intrinsic Factors \\
     \hline

    \end{tabular}
    \caption{Examples of irrelevant matched pairs and the corresponding possible reasons.}
\end{table*}

As previously mentioned, the similarity between sentences does not necessarily indicate their relevance. In this section, we discuss matched pairs that exhibit high similarity between release notes and user reviews but are not labeled as relevant. Generally speaking, there are four possible reasons:

\textbf{Partial matching is a prevalent cause for irrelevant matched pairs, as overlapping keywords in sentences can produce distinct meanings.} In some cases, we found that high similarity between sentences was primarily due to partial keyword overlap, and the absence of other parts of the phrase led to a completely different meaning. For instance, in Table~\ref{tab_irrelevant} (column five), we present an example of an irrelevant matched pair where the release note refers to \textit{Siri support on Apple Watch}, while the user review asks for \textit{Apple Watch support}. While both contain the keyword ``Apple Watch'', they describe completely different features.

\textbf{Semantic matching may result in mismatches, even when the matched pairs share comparable semantics.}  While semantic matching can often improve the hit ratio of matched pairs, it can also lead to mismatches. For instance, in Table~\ref{tab_irrelevant} (column five), Spotify introduces \textit{iPhone X support} in a release note, while an irrelevant review requests an update to \textit{support for iPad Pro}. While both devices are made by Apple and share similar semantics, the high degree of similarity between the two phrases can lead to mismatches.

\textbf{Special symbols and numbers can lead to mismatches, since they may be removed during data preprocessing.} As previously mentioned in Section~\ref{dp}, we performed data preprocessing to remove special symbols and numbers. While this is a basic operation for most natural processing tasks, it can have a significant impact on experimental results. For instance, in Table~\ref{tab_irrelevant}, the release note in row six introduces a feature related to hashtags on Instagram, but we failed to match user reviews because many users used the symbol \# to describe hashtags, which was removed during preprocessing. Similarly, in the release note in row seven, the iOS version number ``12'' was removed during preprocessing, which may have led to the matching failure.

\textbf{Intrinsic factors, such as the lack of pertinent user reviews for certain release notes, can also cause mismatches.} We found that some release notes fail to match relevant user reviews due to a lack of or insufficient matched pairs. While we selected the top 80 matched pairs with the highest similarity for each model in our experiments, some release notes may have far fewer than 80 relevant reviews, or even no relevant reviews at all. Although \textsc{RoseMatcher} can reduce this issue to some extent by taking the intersection of matched pairs to obtaining high-confidence results, there are still cases where mismatches occur with both models. For instance, consider the release note \#8, where both Word2Vec and SBERT suggest ``\textit{Can you add create a list for me and for others}'' as relevant user reviews. Upon manually conducting keyword searches (including synonyms and potential user expressions) for SHEIN's user reviews, we discovered that there were no reviews related to \textit{community activity}. Thus, the two models may have matched some user reviews that are not closely relevant to the release notes.

\subsection{Insights from the performance of \textsc{RoseMatcher}}

Our study is an empirical investigation in which we utilize \textsc{RoseMatcher} to match relevant pairs of release notes and user reviews, shedding light on the role of user reviews in app updates. Our results provide several key insights:

\textbf{BERT-based models have marginal performance gains over Word2Vec, and combining Word2Vec and SBERT in an ensemble approach significantly enhances the matching hit ratio.}  Despite the larger parameter space and superior language encoding abilities of BERT-based models like SBERT, our experiments demonstrate only a slight improvement over Word2Vec. Additionally, BERT-based models require significantly more running time without substantially increasing the hit ratio. An ensemble approach that combines Word2Vec and SBERT yields better hit ratios, demonstrating the benefits of integrating multiple models' strengths. The enhanced performance is primarily attributed to the voting mechanism that bolsters self-consistency~\cite{wang2022self}, along with a comprehensive representation of semantic relationships, increased robustness, diverse feature representations, and a reduced risk of overfitting.

\textbf{The overlap of relevant matched pairs generated by the two models is minimal.} Although the two models have matched a similar number of relevant pairs, the intersection size of the matched pairs given by the two models is much smaller than the size of the union set. This indicates that the two models have different emphases and attentions in similarity matching, and many matched pairs that are only hit by a single model are discarded. However, considering the cost and efficiency of manual annotation and the nature of our empirical research, we believe that this 
treatment
is a reasonable and efficient way to expand the number of researched release notes.

\textbf{The hit ratio of \textsc{RoseMatcher} demonstrates considerable variation across five distinct apps.} The hit ratio of \textsc{RoseMarcher} exhibits a significant variation across five different apps, as depicted in Figure~\ref{fig:hit}. The maximum hit ratio reaches 0.963 (Instagram), whereas some apps (e.g., SHEIN) achieve a meager hit ratio of 0.05. The diverse update habits of different apps could be one of the reasons for such variability. Our study found that SHEIN and Reddit exhibit relatively low hit ratios. Reddit's release notes tend to be excessively detailed and specific, which might not interest a vast number of users, resulting in lower attention and engagement rates. Conversely, SHEIN's updates often lack specific details and remain too general, making it difficult for \textsc{RoseMarcher} to match them with user reviews. It is crucial to note that the release notes analyzed in our study were randomly selected and may not be entirely representative of all release notes for each app. Additionally, the small sample size of release notes might also contribute to the observed variations in hit ratios across different apps. Moreover, another potential reason could be the difference in user demographics among different apps. For instance, some apps may appeal to younger users, who may express themselves differently compared to older users. Therefore, the language model used in \textsc{RoseMarcher} might have more difficulty matching the language used in release notes with user reviews. Furthermore, the nature of the app's updates could be a contributing factor. For example, if an app primarily releases bug fixes and minor improvements, there may be a lack of substantial changes in the release notes, leading to fewer matches with user reviews. Conversely, if an app releases significant new features, there might be more user reviews discussing those features, leading to more matches with the release notes.

\subsection{Bi-directional Communication between Users and Development Teams}

\textbf{Matching results suggest that development teams may consider user reviews to some degree when updating apps.} Although there is no evidence to prove that development teams refer to user reviews for updates, the matching results of \textsc{RoseMatcher} have successfully identified some user reviews with corresponding requests, bug reports, or complaints in many release notes. These matches indicate that development teams pay attention to user feedback, and suggest that software requirements engineering research on user reviews is valuable.

\textbf{User reviews can offer valuable insights for improving app release notes.} Our study indicates that some of the user reviews can potentially be considered as direct feedback on app release notes, which can be targeted and of great value to software iterations. Thus developers could write detailed and easy-to-understand release notes, this will allow users to quickly digest the features or bug fixes brought by new updates. This can possibly lead to high-quality follow-up user feedback, improving the communication between users and development teams.

\textbf{Multiple factors, not solely user reviews, may influence software updates.}  While user reviews can be an important source of feedback for software updates, they may not be the only driver. Our research using \textsc{RoseMatcher} has shown that some release notes fail to match any relevant user reviews, even after performing more in-depth searches. This suggects that software updates may come from internal development teams, project managers, specialized decision or strategy-making departments, or even competitors. Thus, future studies may consider other data sources beyond user reviews to explore the various drivers of software updates.

\textbf{Employing \textsc{RoseMatcher} can assist development teams in utilizing user feedback for app updates more effectively.} \textsc{RoseMatcher} can automatically obtain feedback from users on specific app updates. This enables development teams to quickly identify user sentiments towards a new release, respond rapidly to feedback, and make necessary adjustments. Moreover, \textsc{RoseMatcher} can help search for users who are experiencing bugs or technical issues concerning with certain features, allowing development teams to prioritize and address these concerns efficiently.

\section{Implications}
\label{implication}

\textbf{Implications for Researchers.} Our study has several implications for researchers in the field of software engineering and natural language processing. First, given the rapid advancements in NLP, researchers should consider trying state-of-the-art methods and techniques to further improve the performance of matching and analyzing textual data in software engineering tasks. Second, the analysis of the roles that user reviews play in app updates highlights the importance of user-centric approaches in software development research. By focusing on these roles, researchers can gain valuable insights into user behavior and preferences, which can inform future studies in user-centric software development.

Furthermore, the observed differences in user groups, such as age-related variations in feedback, present interesting research directions for investigating the impact of user demographics on app updates and user satisfaction. Researchers could explore how different user groups interact with and influence app development, tailoring models and techniques to better cater to diverse user needs. Also, our study encourages researchers to positively explore the application of LLMs in the Software Engineering domain. As these models continue to advance, they offer exciting opportunities for improving various software engineerings tasks, such as requirements analysis, user feedback processing, and bug identification. Researchers should investigate the potential benefits and challenges of incorporating such models into software engineering research and practice.

We suggest that future research 
could investigate the effectiveness of classification-based approaches as opposed to simple similarity matching. Our study highlights that similarity does not necessarily equate to relevance, and that relevance matching requires consideration of the specific features and context in which the text is used, instead of similar features. Our study highlights the benefits of combining multiple approaches to improve the accuracy of relevance matching. Specifically, we found that combining clustering of similar user reviews and release notes using different models produced a significant improvement in hit rate. Future research could explore alternative combinations of techniques and methods to further improve the performance of relevance matching in software engineering tasks. Overall, we believe that exploring advanced methods for relevance matching and combining multiple approaches holds significant promise for improving the accuracy and effectiveness of natural language processing in software engineering research and practice.

\textbf{Implications for Development Teams.} Our study highlights the crucial role of user reviews in driving app evolution and identifies eight distinct user roles that contribute valuable feedback to inform app development and maintenance practices. In the following, we discuss the implications of these findings for development teams and how they can leverage this knowledge to enhance their app development process.

\begin{itemize}

    \item \textsc{Feature Requesters} (30.5\%): Development teams should pay close attention to the innovative ideas and suggestions provided by feature requesters. By analyzing and categorizing these requests, developers can create a roadmap for future updates that align with user needs and desires, ultimately leading to increased user engagement, satisfaction, and loyalty.
    \item \textsc{Bug Reporters} (20.2\%): Bug reporters help maintain the app's stability and reliability by identifying technical issues or errors. Developers should prioritize addressing the most critical and widespread issues reported by bug reporters, ensuring a more stable and user-friendly app that fosters trust and positive feedback.
    \item \textsc{Complainers} (20.2\%): Complainers offer valuable insight into areas where the app may be falling short. By addressing these pain points, developers can make targeted improvements that lead to better user satisfaction.
    \item \textsc{Praisers} (12.9\%): Positive reinforcement from praisers can help developers understand which aspects of the app are resonating with users. Developers should capitalize on these strengths and promote successful features more prominently.
    \item \textsc{Dispraisers} (4\%): Dispraisers offer contrasting opinions, pointing out areas where updates or features may not have met expectations. By learning from missteps and addressing these concerns, developers can refine their approach and enhance the app's overall quality.

    \item \textsc{Quality Issue Raisers} (11.4\%): Quality issue raisers focus on specific features of the app and point out potential issues related to quality, functionality, or user experience. Developers should analyze the feedback from quality issue raisers to identify patterns or recurring issues that may require immediate attention, leading to improved functionality and user experience.

    \item \textsc{Subsequent Feature Requesters} (3.8\%): These users provide valuable input by offering suggestions for further improvements or expansions based on an existing feature. Developers should carefully consider subsequent feature requests to identify opportunities to enhance existing features, making them more versatile, user-friendly, or effective.

    \item \textsc{Questioners} (1.7\%): Questioners often raise queries or express confusion about specific aspects of the app. Developers should address the concerns raised by questioners to enhance the overall user experience, making it more intuitive and accessible for a wider audience.
\end{itemize}

By understanding the importance of different user roles, addressing their feedback, and incorporating their unique perspectives into the development process, teams can create a more comprehensive and well-rounded app that caters to the diverse needs and preferences of its user base. This holistic approach to app development ensures that a wide range of user concerns, ideas, and expectations are taken into consideration, resulting in a more engaging, user-friendly, and successful app that resonates with its target audience.

In addition to leveraging the feedback provided by different user roles, development teams may consider incorporating some open-source state-of-the-art Large Language Models (LLMs), such as 
PALM~\cite{chowdhery2022palm}, OPT~\cite{zhang2022opt}, and LLaMA~\cite{touvron2023llama}, into their development process. By using LLMs to automatically analyze and categorize user reviews, developers can effectively extract crucial insights that guide their development decisions, leading to a more efficient, user-focused development process and higher-quality apps that cater to user needs and preferences.

\textbf{Implications for App Store Managers.}
App store managers can also benefit from this study by gaining insights into the factors that influence user satisfaction and engagement with apps. An understanding of how user reviews influence app updates can inform app store policies that promote user-centric development. By encouraging the development teams to pay attention to user reviews, give response promptly and improve app quality, app store managers can promote a more positive app ecosystem that benefits both users and developers. Additionally, the connection and communication between users and development teams are significantly highlighted in our study, thus, app store managers can advocate for more transparent and informative release notes by encouraging the development teams to clearly explain the changes made in each update and how they address user feedback. This would help users understand the improvements made in each update and foster better communication between users and developers.

\textbf{Implications for Users.}
Users can benefit from the findings of this study as it demonstrates that their feedback does indeed have an impact on app updates. First, by understanding the various roles their reviews play in app updates, users can be motivated to write more constructive suggestions, providing valuable input that contributes to the development and improvement of their favorite apps. Second, users can treat app platforms, such as App Store and Google Play, as a reliable bridge between the development teams and them to propose requirements and utilize them effectively, fostering better communication. Lastly, users can and should expect to see more responsive and user-focused app updates as developers become increasingly aware of user feedback. By encouraging the development teams to prioritize user needs, users can contribute to improving the app ecosystem, ultimately leading to a better overall app experience.

\section{Threat to Validity}
\label{tv}
In this section, we elaborate on the threats to the validity of this study, including construct validity, internal validity, external validity, and conclusion validity, by following the guidelines in \cite{wohlin2012experimentation}.

\subsection{Construct Validity}

Construct validity denotes whether the theoretical and conceptual constructs are correctly measured and interpreted. We raise the following two aspects concerning this validity:

First, the relevance between user reviews and release notes does not completely mean that the updates of the development teams come from these reviews or the users give feedback after reading the release notes since we cannot give full assurance only from the semantic similarity and time level. Therefore, this article does not assert that the relationship between the two can represent the communication between developers and users, but we are happy and excited to detect the mutual response between them and conduct further analysis of these matched pairs.

Second, the sources of the app updating by development teams are diverse. Apart from user reviews in the app store, they may originate within the app company, since product managers in mature organizations are mainly responsible to find out what the reasons for the feature requests are, where the user needs and pain points are, and companies often set up special testing departments for bug detection and fixing. Additionally, some social media, such as Twitter, and questionnaires, may also become sources of updates. While some minor updates may not attract users' attention at the same time, those who are willing to give feedback are even less. From this point of view, some matching studies on release notes may become meaningless, which is also the reason for the low hit ratio of the approach, since there may be no relevant user reviews at all.

\subsection{Internal Validity}
Internal validity refers to the degree to which a conclusion concerning the causal effect of one variable (independent or treatment variable) on another variable (dependent variable) based on an experimental study is warranted.

First, release notes and user reviews are two different natural languages, where large differences can be detected in their expression methods, tone, and sentiment. Even if the word embedding models try to eliminate the language gap, it will be difficult to avoid the errors brought by the algorithm and the calculation deviation caused by the sentiment difference between them.

Second, we considered verbals, nouns, and adjectives (giving the same weight) when calculating sentence-level similarity, but we observed that release notes, being natural language, have a lot of interference information, which may lead to the inaccurate matching of user reviews. The research of how to identify these interfering information and automatically remove them, or give higher weight to key information in the sentence vector may improve the accuracy of the matching algorithm.

\subsection{External Validity}

External validity concerns the extent to which the results of a study can be generalized to and across other situations, people, settings, and measures. The main external threat to validity is concerned with the generalizability \cite{generalizability} of our findings. Our study focuses on analyzing the 5 apps from the Apple App Store over a period of five years. Furthermore, our study focuses on the apps that are the most popular in the App Store. Our findings may not hold for non-popular apps. Similarly, the studied 5 apps cannot cover the 26 app categories, and it is worthwhile to include other categories of apps to increase the external validity. Further studies should investigate more apps over a more extended period to understand how our findings apply to other types of apps, such as apps in Google Play Store and non-free apps.

\subsection{Conclusion Validity}
Threats to conclusion validity are concerned with the relationship between the treatment and the outcome. In our study, a relevant threat concerns the update version which users post reviews for. It is possible that the reviews posted by users are not for the latest version, because it is highly likely that users will have a lag in version updates: since using older versions of apps will not affect the experience, for those who do not have automatic updates enabled, their reviews will have a certain lag. In this way, the following extreme case may exist, where the user is using 1st version of the app, but he does not update the app timely, the reviews he posted are between the 4th and the 5th version, which will cause some bias to our result analysis.

\section{Related Work}
\label{rw}

\subsection{Extracting User Requirements from User Reviews}
Data-driven requirements engineering (RE) is increasingly active in RE research, which usually takes different types of user feedback of software as the most important data source. In these studies, user reviews are reported as the most popular data source, since user reviews contain unique insights from users based on their experiences, such as problem reports and inquiries~\cite{Multilingual2019}, which results in user feedback having significant value to software engineers and requirements managers~\cite{Data-Driven}. 

In recent years, many researchers focused on requirements extraction from a large number of user reviews for app evolution. For example, Chen et al. proposed AR-Miner~\cite{ARMiner} to extract informative user reviews by filtering noisy and irrelevant ones, group the informative reviews automatically using topic modeling, prioritize the informative reviews by an effective review ranking scheme, and finally present the groups of most `informative' reviews via an intuitive visualization approach.
Casper~\cite{Casper} was presented by Guo et al. as a method to extract and synthesize user-reported mini-stories regarding app problems from reviews.
Miroslav et al.~\cite{KeywordAssistedTopicModels} applied domain-specific analysis and keyword-assisted approach to extract useful information from app user reviews. These studies reported that their proposed keyword-assisted topic modeling approach was able to significantly outperform LDA on both intrinsic and extrinsic measures of topic cohesiveness. Sorbo et al.~\cite{di2016would} developed an approach called SURF (Summarizer of User Reviews Feedback) which utilizes a conceptual model to capture user needs and summarization techniques to generate a condensed agenda of recommended software changes from thousands of user reviews.

In addition, several studies have explored in depth the types of user reviews based on their content, while also factoring in user ratings into the analysis. 
For instance, Pagano and Maalej \cite{UserFeedback} reported 17 topics from app reviews. They found that 96.4\% of reviews with one-star ratings include the topics of shortcomings or bug reports, and these reviews can be further mined to get requirements.
To alleviate app developers' effort to identify the user reviews that matter to be addressed for receiving higher star ratings, Noei et al. \cite{TooManyUserReviews} considered the key topics of user reviews. It was found that considering the key topics in the next version shares a significant correlation with the increase in star ratings. 
Sorbo et al. \cite{ Criticality} combined the usage of app ratings and user reviews. They found that user comments reporting bugs are negatively correlated with the rating, while reviews reporting feature requests do not. 

These studies aimed to provide multifaceted approaches for requirements extraction, category classification and requirement prioritization based on user reviews. However, little is known about whether these user requirements or bug reports are adopted or resolved by the development teams and written in the app release notes, that is, the value and role of user reviews in app updates remain unclear.
Compared to these studies, our study first introduces \textsc{RoseMatcher} to detect the relevance between user reviews and app release notes to fill the gap. Subsequently, by analyzing the relevance, we delve into the roles that user reviews play in app updates. 

\subsection{Analyzing app evolution with release notes}
Release note provides valuable and important changes of an app to the users. Descriptive and effective release notes will help app vendors re-engage users, build excitement around the entire product, and extend app vendors' reach to new audiences~\cite{Appcues}.

Prior research on app release notes focuses on the content and the evolution principles. 
For example, Wang et al. \cite{Chong} analyzed release notes pertaining to 120 apps in the Apple App Store from the perspectives of FR (Functional Requirement) and NFR (Non-Functional Requirements). They found that the majority of the changes in fact refer to NFR and developers seem to be busy with improving the quality aspects of their apps, and relatively fewer changes refer to FR. 
Hassan et al. \cite{EmergencyUpdates, BadUpdates} studied the characteristics of emergency updates and bad updates, which shows that release notes are essential tools for app development teams to announce the resolution of emergency issues and the implementation of the requested features.
The survey research by Nayebi et al. \cite{ReleasePractices} focused on the ways in which mobile app developers organize their releases and the release strategies they employ. The authors found that half of the developers participating in the survey had a clear strategy for their app releases.
Yang et al.~\cite{AnEmpirical} conducted an in-depth analysis of the release notes practice in the Google Play Store and identified six patterns of release notes. They also noticed that the shifting of patterns mainly occurs when developers shift from short to long and rarely updated to frequently updated release notes.

Some researchers explored the impact of the contents of the release notes or the frequency of app updates on user ratings. For instance, McIlroy et al.~\cite{FreshApps} explored the update frequency of the top 10,713 mobile apps across 30 mobile app categories. Their results indicate that 14\% of the apps are updated frequently, while 45\% of these frequently-updated apps do not provide the users with any information about the rationale for the new updates. Plus, frequently-updated apps are highly ranked by users.

Different from these existing work, this paper takes an in-depth analysis on the combination and relevance of two different natural languages, i.e., user reviews and release notes.

\subsection{Tracking User Reviews to Support App Evolution}
Some researchers have tried to track user reviews of mobile applications to support app evolution. Wang et al. ~\cite{apsec2021} tried to match user reviews with app release notes in Spotify and manually labeled types of the user reviews. Häring et al.~\cite{DeepMatcher} proposed an automatic approach DeepMatcher to extract problem reports from app reviews submitted by users, and then identify matching bug reports in an issue tracker used by the development team, which can help developers identify bugs earlier, enhance bug reports with user feedback, and eventually lead to more precise ways to detect duplicate or similar bugs. 
Palomba et al. proposed an approach named CRISTAL~\cite{CRISTAL} to trace informative crowd reviews into code changes, in order to monitor the extent to which developers accommodate crowd requests and follow-up user reactions as reflected in their ratings. Their results indicate that developers implementing user reviews are rewarded in terms of ratings. Zhou et al. ~\cite{zhou2020user} introduced an automated approach called RISING, which uses domain-specific constraints and semi-supervised learning to cluster and classify user reviews and identify potential change files, to link user reviews to potential files to be changed (such as source code) in mobile app development. Our study differs from them in the following aspects. On the one hand, they narrowed down their exploration to bug reports in issues trackers and source code changes from a few open-source projects. However, the bug fixes and code changes made by developers are invisible to the vast majority of users, while the release notes serve as a bridge for app vendors to notify users of app newest updates. The mutual pushing of app release notes and user reviews makes the app store an interactive platform, which indicates that our match will be double-oriented. On the other hand, our paper empirically studies the roles of user reviews in app updates according to their relevance with release notes, instead of providing developers with the functionality to identify bug reports or new feature requests. 

Moreover, several studies have also attempted to explore the relationship between release notes and user reviews. Villarroel et al.~\cite{CLAP} proposed an approach called CLAP (Crowd Listener for release Planning) to categorize user reviews, cluster together related reviews and prioritize the clusters of reviews to suggest app release planning. In a follow-up study, Scalabrino et al.~\cite{ListeningToCrowd} improved CLAP for app developers to plan for the next release by selecting the most important raised complaints in a particular issue type. For evaluation, they used release notes as the basis, aiming to verify whether the issues clustered and prioritized by CLAP were resolved by the developers in subsequent updates. 
Unlike these authors' studies~\cite{CLAP, ListeningToCrowd}, our work is based on the release notes, trying to identify the most relevant user reviews to determine the role of user reviews in app evolution and updates.

In addition, Saidani et al.~\cite{TrackBadUpdate} proposed AppTracker, a novel approach to automatically track bad release updates in Android applications. They judge whether an update is good or not from the level of positive and negative reviews given by users. While their work focused on the degree to which users respond positively or negatively to developers, our study does not only focus on the sentiment of user reviews responding to release updates, but we also analyze the roles played by user reviews on the release notes, including but not limited to whether user reviews may be a trigger for the delivering of new release notes, and whether the new release notes may lead to new reports of quality problems or new related feature request.

\section{Conclusions and Future Work}
\label{cfw}

In this paper, we introduce \textsc{RoseMatcher}, an approach to automatically match colloquially-written user reviews with technically-written release notes, and identify the relevant matched pairs, which can not only address the language gap between the two natural languages but also greatly improve the hit ratio of the matched pairs. Leveraging \textsc{RoseMatcher} on 72 release notes from 6 different apps, we explored these relevant matched pairs in depth: from the perspective of the content, we defined eight roles of user reviews in app update. We found that the feature requests, bug reports, and complaints raised by users will be adopted by the development teams and solved in the update according to the release notes pushed regularly. Meanwhile, users will also give feedback to the app release notes, including praise, criticism, bug reports, questions, and novel ideas related to the update. From the perspective of time, we found that the relevant reviews of release notes tend to concentrate on a short time before and after the release notes are released, but there are also plenty of matched pair cases with long intervals, some even as long as three years, with an average interval of one year. 

In our future work, we plan to expand our dataset to improve the generalizability of our findings, ensuring that the results are applicable across a wider range of applications and user demographics. Additionally, examining the impact of responding to user reviews and giving feedback to the release notes on the app ratings would deepen our understanding of the interaction between users and development teams and shed light on their potential effects. We also intend to explore the state-of-the-art models to further enhance the predictive capabilities and accuracy for review and release note pair matching. A promising direction involves incorporating Large Language Models (such as LLaMA~\cite{touvron2023llama}) into the analysis process, with the goal of automating the entire analysis pipeline and potentially uncovering more nuanced insights from user reviews. Lastly, we intend to investigate the matching differences of various apps based on their targeted age groups, categories, and other factors, which will help to identify unique patterns and trends within specific user groups of mobile apps.

\section*{Acknowledgements}
This work is supported by the National Natural Science Foundation of China under Grant Nos. 61702378, 61972292, 62032016, and 62172311.

\section*{CRediT authorship contribution statement}
\textbf{Tianyang Liu:} Conceptualization, Methodology, Software, Validation, Formal analysis, Investigation, Resources, Data Curation, Writing - Original Draft, Writing - Review \& Editing, Visualization. \textbf{Chong Wang:} Conceptualization, Methodology, Formal analysis, Investigation, Writing - Review \& Editing, Supervision. \textbf{Kung Huang:} Validation, Data Curation, Visualization. \textbf{Peng Liang:} Conceptualization, Methodology, Formal analysis, Writing - Review \& Editing, Supervision. \textbf{Beiqi Zhang:} Formal analysis, Writing - Review \& Editing. \textbf{Maya Daneva:} Formal analysis, Writing - Review \& Editing. \textbf{Marten van Sinderen: }Formal analysis, Writing - Review \& Editing.



\balance

\bibliographystyle{elsarticle-num} 
\bibliography{elsarticle}





\end{document}